
\input phyzzx
\Pubnum={\vbox{\hbox{CERN-TH.6976/93}\hbox{hep-th/9310046}}}
\pubnum={\vbox{\hbox{CERN-TH.6976/93}\hbox{hep-th/9310046}}}
\date={August, 1993}
\pubtype={}

\def\sl{ SL(2,${\rm I}\!{\rm R}$)}
\def\slU{SL(2,${\rm I}\!{\rm R}$)/U(1)}

\titlepage

\title{INTERACTIONS IN THE \slU \break BLACK HOLE
BACKGROUND}

\vskip 1.0cm

\author{ K. Becker\foot{$^{\dagger}$ Permanent Address:
Universit\"at Bonn, Physikalisches Institut, Nussallee 1,
W-5300 Bonn 1, Germany,} \break and \break M.
Becker$^{\dagger}$}\address{Theory Division, CERN\break
CH-1211 Geneva 23, Switzerland}.

\abstract{We calculate two- and three-point tachyon
amplitudes
 of the
{\slU} two- dimensional Euclidean black hole for spherical
topologies in the continuum approach proposed by Bershadsky
and Kutasov. We find an interesting relation to the
tachyon scattering amplitudes of standard
non-critical string theory.}

\endpage
\pagenumber=1

\def\np{{\it Nucl. Phys. }}

\def\pl{{\it Phys. Lett. }}
\def\cmm{{\it Commun. Math. Phys. }}
\def\ijmp{{\it Int. J. Mod. Phys. }}
\def\mpl{{\it Mod. Phys. Lett. }}
\def\prl{{\it Phys. Rev. Lett. }}
\def\p{\partial}

\def\b{\beta}
\def\g{\gamma}
\def\ap{\alpha_+}
\def\R{{\rm I\!R }}

\def\a{\alpha}
\def\G{\Gamma }
\def\D{\Delta}
\def\e{\varepsilon}
\def\IC{{\ \hbox{{\rm I}\kern-.6em\hbox{\bf C}}}}
\def\IZ{{\hbox{{\rm Z}\kern-.4em\hbox{\rm Z}}}}

\def\rhaw{S. Hawking,``Particle Creation by Black Holes'',
\cmm {\bf 43} (1975) 199.}

\def\rWitt{E. Witten, ``Two-dimensional String Theory and
Black Holes'', Lecture given at the Conference on Topics in
 Quantum Gravity, Cincinnati, OH, Apr 3-4, 1992,
hep-th/9206069.}

\def\rWit{E. Witten, ``String Theory and Black Holes'',
{\it Phys. Rev.} {\bf D 44} (1991) 44. }

\def\rdvv{R. Dijkgraaf, E. Verlinde and H. Verlinde,
``String Propagation in Black Hole Geometry'', \np
{ \bf B 371} (1992) 269.}

\def\rbk{M. Bershadsky and D. Kutasov, ``Comment on Gauged
WZW Theory'', {\it Phys. Lett} {\bf B 266} (1991) 345.}

\def\rms{E. Martinec and S. L. Shatashvili, ``Black Hole
Physics and Liouville Theory'', \np {\bf B368} (1992) 368.}

\def\reky{T. Eguchi, H. Kanno and S. Yang, ``$W_{\infty}$
Algebra in Two-Dimensional Black Hole'', \pl {\bf B 298}
(1993) 73.}

\def\rgm{P. Ginsparg and G. Moore, ``Lectures in 2D Gravity
and 2D String Theory'', Yale Preprint YCTP-P23-92,
LA-UR-92-3479, hep-th/9304011.}

\def\rdn{J. Distler and P. Nelson, ``New Discrete States of
Strings Near a Black Hole'', \np {\bf B 374} (1992) 123.}

\def\rw{M. Wakimoto, ``Fock Representations of the Affine
Lie Algebra ${A_1}^{(1)}$'', \cmm {\bf 104} (1986) 605.}

\def\rd{V. S. Dotsenko, ``The Free Field Representation of
the SU(2) CFT'', \np {\bf B 338} (1990) 747; ``Solving the
SU(2) CFT with the Wakimoto Free Field Representation'', \np
{\bf B358} (1991) 547.}

\def\rgmom{A. Gerasimov, A. Morozov, M. Olshanetsky and A.
Marshakov, ``Wess-Zumino-Witten Model as a Theory of Free
Fields'', \ijmp {\bf A 5} (1990) 2495.}

\def\rdfk{P. Di Francesco and D. Kutasov, ``World Sheet and
Space Time Physics in Two Dimensional (Super-)String
Theory'', \np {\bf B375} (1992) 119; ``Correlation Functions
in 2-D String Theory'', \pl {\bf B 261} (1991) 385.}

\def\rdtwo{V. S. Dotsenko, ``Correlation Functions of Local
Operators in 2-D Gravity Coupled to Minimal Matter'', Summer
School Carg\`ese 1991, hep-th/9110030; ``Three Point
Correlation Functions of Minimal CFT Coupled to 2-D Gravity,
\mpl {\bf A 6} (1991) 3601.}

\def\rgl{M. Goulian and M. Li, ``Correlation Functions in
Liouville Theory'',
\prl {\bf 66} (1991) 2051.}

\def\re{T. Eguchi, ``$c=1$ Liouville Theory Perturbed by the
Black Hole Mass Operator'', Tokyo Preprint UT 650, July
1993, hep-th/9307185.}

\def\rjy{A. Jevicki and T. Yoneya, ``A Deformed Matrix Model
for the Black Hole Background in Two-Dimensional String
Theory'', Santa Barbara Preprint NSF-ITP-93-67,
hep-th/9305109; K. Demeterfi and J. P. Rodrigues, ``States
and Quantum Effects in the Collective Field Theory of the
Deformed Matrix Model'' Princeton Preprint PUPT-1407,
hep-th/9306141; U. Danielsson, ``A Matrix Model Black
Hole'', CERN-TH-6916, hep-th/9306063; K. Demeterfi, I.
Klebanov and J. P. Rodrigues, ``The $S$ Matrix of the
Deformed $c=1$ Matrix Model'', Princeton Preprint PUPT-1416,
hep/th-9308036.}

\def\rbih{W. J. Holman and L. C. Biedenharn, ``A general
Study of the Wigner Coefficients of SU(1,1)'', {\it Ann.
Phys} {\bf 47} (1968) 205;
V. Bargman, ``Irreducible unitary representations of the
Lorentz group'', {\it Ann. Math.} {\bf 48} (1947) 568. }

\def\rdfa{V. S. Dotsenko and V. A. Fateev, ``Four point
correlation function and the operator algebra in the 2D
conformal invariant theories with central charge $c<1$'',
\np {\bf B 251} (1985) 691; V. S. Dotsenko and V. A. Fateev,
``Conformal algebra and multipoint correlation functions in
2D statistical models'', \np {\bf B 240} (1984) 312.}

\def\rfms{D. Friedan, E. Martinec and S. Shenker,
``Conformal Invariance, Supersymmetry and String Theory'',
\np {\bf B 271} (1986) 93.}

\def\rlz{B. H. Lian and G. J. Zuckerman, ``2D Gravity with
$c=1$ Matter'', \pl {\bf B 266}(1991) 21. }

\def\rsei{N. Seiberg, ``Notes on Quantum Liouville Theory
and Quantum Gravity'', in Common Trends in Mathematics and
Quantum Field Theory, Proc. of the 1990 Yukawa International
Seminar, {\it Prog. Theor. Phys. Suppl.} {\bf 102} (1990)
319, and in Random Surfaces and Quantum Gravity, proceedings
of 1990 Carg\'ese Workshop, edited by O. Alvarez, E.
Marinari and P. Windey.}

\def\rcl{S. Chaudhuri and J. Lykken, ``String Theory, Black
Holes and {\sl} Current Algebra'', \np {\bf B 396} (1993)
270.}

\def\remn{J. Ellis, N. E. Mavromatos and D. V. Nanopoulos,
``Quantum Coherence and Two-Dimensional Black Holes'', \pl
{\bf B 267}(1991) 465. }

\def\rmo{N. Marcus and Y. Oz, ``The Spectrum of 2-D Black
Hole or does the 2-D Black Hole have Tachyonic $W$ Hair?'',
Tel Aviv University preprint, TAVP-2046-93, hep-th/9305003.}

\def\rmoore{G. Moore, ``Finite in All Directions'', Yale
Preprint YCTP-P12-93, hep-th/9305139.}

\def\rvm{S. Mukhi and C. Vafa, ``Two Dimensional Black Hole
as a Topological Coset of $c=1$ String Theory'', Harvard
preprint HUTP-93-A002 , hep-th/9301083.}

\def\rkutasov{D. Kutasov, ``Geometry On the Space of
Conformal Field Theories and Contact Terms'', {\pl} {\bf B
220} (1989) 153.}

\def\rgs{M. B. Green and N. Seiberg, ``Contact Interactions
in Superstring Theory'', {\np} {\bf B 299} (1988) 559.}

\def\reder{A. Erd\'elyi, ``Higher Trascendental Functions'',
p. 188-189, New York: McGraw-Hill, 1953.}

\def\rvil{N. Ja. Vilenkin and A. U. Klimyk, ``Representation
of Lie Groups and Special Function'', Mathematics and Its
Applications, Kluwer Academic Publishers, Dordrecht, 1992.}

\def\Ai{$$
S_{WZW}(g)={k \over 8\pi}\int_{\Sigma}d^2x\sqrt{h} h^{ij}
tr( g^{-1} \p_i g g^{-1} \p_j g)+ik\Gamma(g),
\eqn\Ai
$$}

\def\Aii{$$
\Gamma (g)={1\over 12 \pi}\int_B d^3 y \varepsilon^{abc}
 tr(g^{-1}\p_a g
g^{-1}\p_b g g^{-1} \p_c g),
\eqn\Aii
$$}

\def\Aiii{$$
S_{WZW}(g,A)=S_{WZW}(g)+{k\over 2\pi} \int d^2z({\bar A}
tr(Gg^{-1}\p g)+A tr (G {\bar \p} g g^{-1})+A{\bar A}
(-2+tr( GgGg^{-1})))
\eqn\Aiii
$$}

\def\Aiv{$$
S_{WZW}^{gf}=S_{WZW}(g)+S(\Phi)+S(b,c),
\eqn\Aiv
$$}

\def\Av{$$
S(\Phi)=-{k\over 4\pi}\int d^2z \p \Phi {\bar \p}\Phi,
\eqn\Av
$$}

\def\Avi{$$
S(b,c)=\int d^2 (b{\bar \p} c +{\bar b}\p {\bar c}).
\eqn\Avi
$$}

\def\Avii{$$
J=J^a t^a =-{k\over 2} \p g g^{-1} \qquad , \qquad
{\bar J}={\bar J}^a t^a =-{k\over 2} g^{-1}
{\bar \p} g ,
\eqn\Avii
$$}

\def\Aviii{$$
J_{+}(z)J_{-}(w)={k\over (z-w)^2}-{2J_3(w)\over (z-w)}
+\dots
$$
$$
J_3(z)J_{\pm}={\pm}{J_{\pm}(w)
\over (z-w)}+\dots
$$
$$
J_3(z)J_3(w)=-{{k\over 2}\over (z-w)^2}+\dots.
\eqn\Aviii
$$}

\def\Bi{$$
\langle \gamma(z) \beta(w) \rangle =-{1\over (z-w)}
\qquad , \qquad
\langle \p \phi(z) \p \phi(w) \rangle =-{1\over (z-w)^2},
\eqn\Bi
$$}

\def\Bii{$$
J_+(z)=\b(z)\qquad , \qquad
J_3(z)=-\b(z)\g(z)-{\ap\over 2}\p \phi(z)
$$
$$
J_-(z)=\b(z)\g^2(z)+\ap\g(z) \p\phi(z)+k\p\g(z).
\eqn\Bii
$$}

\def\Biii{$$
T_{{\rm SL(2,I\!R)}}=-{\Delta\over (k-2)}
\qquad , \qquad
\Delta=-{1\over 2}: J_+J_-+J_-J_+:+:J_3J_3:,
\eqn\Biii
$$}

\def\Biv{$$
T_{{\rm SL(2,I\!R)}}=\b\p\g-{1\over 2}(\p \phi)^2-
{1\over \ap}\p^2\phi.
\eqn\Biv
$$}

\def\Bv{$$
c={3k\over k-2}.
\eqn\Bv
$$}

\def\Bvi{$$
\Delta_0 |j,m\rangle =j(j+1)|j,m\rangle \qquad,\qquad
J_0^3|j,m\rangle =m|j,m\rangle ,
\eqn\Bvi
$$}

\def\Bvii{$$
J_n^{\pm}|j,m\rangle=J_n^3|j,m\rangle =0\qquad
{\rm for}\quad n>0,
\eqn\Bvii
$$}

\def\Bviii{$$\eqalign{
J_0^+|j,m\rangle & =(-m+j)|j,m+1\rangle\cr
& \cr
J_0^{-}|j,m\rangle & =(-m-j)|j,m-1\rangle .\cr}
\eqn\Bviii
$$}

\def\Bix{$$
\eqalign{
&|{\rm HWS\; module }\rangle  =|j,m\rangle
\qquad m=j,j-1,\dots \cr
&|{\rm LWS\; module }\rangle  =|j,m\rangle
\qquad m=-j,-j+1,\dots \cr }
\eqn\Bix
$$}

\def\Bx{$$
(j_1, m_1|j_2\; m_2)=\delta^{j_1 j_2}\delta^{m_1 m_2}.
\eqn\Bx
$$}

\def\Bxi{$$
J_0^+ |j,m) =(j-m){N(j,m)\over N(j,m+1)}|j, m+1)
$$
$$
J_0^-|j,m) =-(j+m){N(j,m)\over N(j,m-1)}|j,m-1).
\eqn\Bxi
$$}

\def\Bxii{$$
(j,m+1|J_0^+ | j,m)^{\ast}=(j,m|J_0^-|j,m+1).
\eqn\Bxii
$$}

\def\Bxiii{$$
\left| (j-m){N(j,m)\over N(j,m+1)}\right|^2 =-(j+m+1)(j-m).
\eqn\Bxiii
$$}

\def\Bxiv{$$
|j,m) ={1 \over
\sqrt{\Gamma(j+m+1)\Gamma(j-m+1)}}|j,m\rangle.
\eqn\Bxiv
$$}

\def\Bxv{$$
J_0^+|j,m)=\sqrt{(j-m)(j+m+1)}|j,m+1)
$$
$$
J_0^-|j,m)=\sqrt{(j+m)(j-m+1)} |j,m-1)
$$
$$
J_0^3|j,m)=m|j,m)
\eqn\Bxv
$$}

\def\Bxva{$$
|j,m\rangle=\sqrt{\G(j+m+1)\G(j-m+1)}|j,m).
\eqn\Bxva
$$}

\def\Bxvi{$$
{\cal V}_{j\; m}(z)=:\g^{j-m}(z) e^{{2\over \ap} j\phi(z)}:,
\eqn\Bxvi
$$}

\def\Bxvii{$$
T(z){\cal V}_{j\;m}(w)={h_{j,m}\over (z-w)^2}
{\cal V}_{j\; m}(w) +{1\over (z-w)}\p
{\cal V}_{j\;m}(w)+\dots
$$
$$
J^a(z){\cal V}_{j\;m}(w)={t^a_{(j,m)}\over (z-w)}
{\cal V}_{j\;m}(w)+\dots
\eqn\Bxvii
$$}

\def\Bxviii{$$
h_{j,m}=-{j(j+1)\over k-2}.
\eqn\Bxviii
$$}

\def\Bxix{$$
S={1\over 2\pi}\int \p \phi{\bar \p}\phi -{2\over \ap}
R^{(2)} \phi +\b{\bar \p}\g +{\bar \b}\p{\bar \g}.
\eqn\Bxix
$$}

\def\Bxx{$$
\langle T(z){\cal V}_1(z_1,{\bar z}_1)\dots
{\cal V}_N(z_N,{\bar z}_N)\rangle =
\sum_{i=1}^N \left( {h_i\over (z-z_i)^2}+
{1\over (z-z_i)}{\p \over \p z_i}\right)
\langle {\cal V}_1(z_1,{\bar z}_1)\dots
{\cal V}_N(z_N,{\bar z}_N)\rangle
$$
$$
\langle J^a(z) {\cal V}_1(z_1,{\bar z}_1)
\dots {\cal V}_N(z_N,{\bar z}_N)\rangle =
\sum_{i=1}^N {t^a_i\over (z-z_i)}
\langle {\cal V}_1(z_1,{\bar z}_1)\dots
{\cal V}_N(z_N,{\bar z}_N)\rangle .
\eqn\Bxx
$$}

\def\Bxxi{$$
{\cal Q}=\int d^2 z J(z,{\bar z})
\qquad , \qquad
J(z,{\bar z})=\b(z){\bar \b}({\bar z})
e^{-{2\over \ap}\phi(z,{\bar z})} .
\eqn\Bxxi
$$}

\def\Bxxii{$$
J_+(z)J(w,{\bar w})\sim reg.\qquad , \qquad
J_3(z)J(w,{\bar w})\sim reg.
$$
$$
J_-(z)J(w,{\bar w})\sim {\p \over \p w}\left(
{e^{-{2\over \ap}
\phi(w,{\bar w})}\over (z-w)}\right).
\eqn\Bxxii
$$}

\def\Bxxiii{$$
S={1\over 2\pi}\int \p \phi{\bar \p}\phi -{2\over \ap}
 R^{(2)} \phi +\b{\bar \p}\g +{\bar \b}\p{\bar \g}
+2\pi M \b {\bar \b} e^{-{2\over \ap}\phi}.
\eqn\Bxxiii
$$}

\def\Bxxiv{$$
\b=-i\p v e^{iv-u} \qquad , \qquad \g=e^{u-iv},
\eqn\Bxxiv
$$}

\def\Bxxv{$$
\langle u(z) u(w)\rangle =\langle v(z) v(w)
\rangle=-\log(z-w).
\eqn\Bxxv
$$}

\def\Ci{$$
\langle \p X(z)\p X(w)\rangle =-{1\over (z-w)^2}
\qquad , \qquad
\langle C(z)B(w)\rangle ={1\over (z-w)}.
\eqn\Ci
$$}

\def\Cii{$$
Q^{U(1)}=\oint C(z)\Bigl(J^3-i\sqrt{{k\over 2}}
\p X\Bigr) dz .
\eqn\Cii
$$}

\def\Ciii{$$
{\cal V}_{j \; m}=:\g^{j-m} e^{{2\over \ap} j\phi}
e^{im\sqrt{{2\over k}}X}:.
\eqn\Ciii
$$}

\def\Civ{$$
h_{j,m}=-{j(j+1)\over k-2}+{m^2\over k},
\eqn\Civ
$$}

\def\Cv{$$
T_{{\rm SL(2,I\!R)/U(1)}}=\b \p \g -{1\over 2}
(\p \phi)^2
-{1\over \ap}\p^2\phi -{1\over 2}(\p X)^2-B\p C.
\eqn\Cv
$$}

\def\Cvi{$$
c={3k\over k-2}-1
\eqn\Cvi
$$}

\def\Cvii{$$
T_{gh}=-2b\p c-\p b c.
\eqn\Cvii
$$}

\def\cv{$$
T_{{\rm SL(2,I\!\R)/U(1)}}=\b \p \g -{1\over 2}
(\p \phi)^2-{1\over \ap}\p^2\phi -{1\over 2}
(\p X)^2-B\p C-2b\p c-\p b c.
\eqn\cv
$$}

\def\Cviii{$$
Q^{Diff}=\oint c(z)\Bigl(T_{{\rm SL(2,\R)/U(1)}}+
{1\over 2}T_{gh}\Bigr),
\eqn\Cviii
$$}

\def\Cix{$$
(Q^{Diff})^2=0\qquad, \qquad \{ Q^{U(1)},
 Q^{Diff}\}=0.
\eqn\Cix
$$}

\def\Cixa{$$
-{j(j+1)\over k-2}+{m^2\over k}=1,
\eqn\Cixa
$$}

\def\Cixb{$$
2j+1=\pm {2\over 3}m.
\eqn\Cixb
$$}

\def\Cx{$$\eqalign{
{\widetilde {\cal D}}^{\pm}&: \hskip 5mm m=\pm
{3(2s-4r-1)\over 8}\qquad,\qquad
j={2s+4r-5\over 8}\cr
&\cr
{\cal D}^{\mp}&: \hskip 5mm m=\pm {3(s-2r+1)\over 4}
\qquad \hskip 2mm,\qquad
j={s+2r-3\over 4}\cr
&\cr
{\cal C}&: \hskip 5mm m= {3(s-r)\over 2}
\hskip 14mm\qquad,\qquad
j={s+r-1\over 2},\cr}
\eqn\Cx
$$}

\def\Cxiii{$$
(L_0-{\bar L}_0)|j,m,{\bar m}\rangle =0\qquad,\qquad
(L_0+{\bar L}_0-2)|j,m,{\bar m}\rangle =0,
\eqn\Cxiii
$$}

\def\Cxiv{$$
L_0|j,m,{\bar m}\rangle={\bar L}_0|j,m,{\bar m}
\rangle=|j,m,{\bar m}\rangle,
\eqn\Cxiv
$$}

\def\cxiv{$$
-{j(j+1)\over k-2}+{m^2\over k}=1,
\eqn\cxiv
$$}

\def\Cxv{$$
2j+1=\pm {2\over 3}m.
\eqn\Cxv
$$}

\def\Cxvi{$$
m={1\over 2}(n_1+n_2k)\qquad,\qquad {\bar m}=
-{1\over 2}(n_1-n_2k)\qquad,\qquad
n_1,n_2\in \IZ.
\eqn\Cxvi
$$}

\def\dia{$$
{\cal A}^{j_1.\,.\,.\,.\,j_N}_{m_1.\,.\,. m_N}
=\langle {\cal V}_{j_1\; m_1} \dots {\cal V}_{j_N\; m_N}
\left(\int d^2z \b{\bar \b} e^{-{2\over \ap}\phi}
\right)^s\rangle_{M=0}
\eqn\dia
$$}

\def\di{$$
s=\sum_{i=1}^N j_i+1
\eqn\di
$$}

\def\dib{$$
\sum_{i=1}^N m_i=0.
\eqn\dib
$$}

\def\dii{$$
\# \b -\# \g =1.
\eqn\dii
$$}

\def\diii{$$
\langle \prod_{i=1}^N {\cal V}_{{j_i}\;{m_i}}
\rangle=M^s\G(-s)
\langle \prod_{i=1}^N {\widetilde
{\cal V}_{{j_i}\;{m_i}}}\left( \int \b{\bar \b}
 e^{-{2\over \ap}
{\widetilde \phi}}d^2z \right)^s\rangle_{M=0},
\eqn\diii
$$}

\def\div{$$
\widetilde{\cal A}^{j_1\;j_2\;\;\;j_3}_{j_1\;m_2\;m_3}=
\int \prod_{i=1}^s d^2z_i
\langle e^{{2\over \ap} j_1
\widetilde{\phi}(0)}e^{{2\over \ap}
 j_2 \widetilde{\phi}(1)}
e^{{2\over \ap} j_3 \widetilde{\phi}(\infty)}
 e^{-{2\over \ap} \widetilde{\phi}
(z_i,{\bar z_i})}\rangle
$$
$$
\langle \g^{j_2-m_2}(1)\g^{j_3-m_3}
(\infty)\b(z_i)\rangle
\langle {\bar \g}^{j_2-m_2}(1)
{\bar\g}^{j_3-m_3}(\infty){\bar \b} ({\bar z}_i)\rangle.
\eqn\div
$$}

\def\dv{$$
{\cal P}=\prod_{i=1}^s(1-z_i)^{m_2-j_2}
\prod_{i<j}(z_i-z_j),
\eqn\dv
$$}

\def\dvi{$$
\widetilde{\cal A}^{j_1\;j_2\;\;\;j_3}_{j_1\;m_2\;m_3}=
\int \prod_{i=1}^s d^2 z_i |z_i|^{-4\rho j_1}
|1-z_i|^{-4\rho j_2}\prod_{i<j}
|z_i-z_j|^{4\rho}
{\cal P}^{-1}{\p^s{\cal P}\over \p z_1\dots \p z_s}
{\bar {\cal P}}^{-1}
{\p^s{\bar {\cal P}}\over \p {\bar z_1}
\dots \p {\bar z_s}}
\eqn\dvi
$$}

\def\dvii{$$
{\cal P}^{-1}{\p^s{\cal P}\over \p z_1 \dots \p z_s}
={\Gamma(-j_2+m_2+s)\over \Gamma(-j_2+m_2)}
\prod_{i=1}^s(1-z_i)^{-1}.
\eqn\dvii
$$}

\def\dviii{$$
\prod_{i=1}^s(z_i-z_j)=\sum_{\sigma(p(1),\dots,p(s))}
 sign(p) z_{p(1)}^0\dots
z_{p(s)}^{s-1},
\eqn\dviii
$$}

\def\dix{$$
{\cal A}^{j_1\;j_2\;\;\;j_3}_{j_1\;m_2\;m_3}=
(-)^s\D(j_2-m_2+1) \D (j_3-m_3+1)
 {\cal I}(j_1,j_2,j_3,k).
\eqn\dix
$$}

\def\dx{$$
{\cal I}(j_1,j_2,j_3,k)=M^s
\G(-s)\int \prod_{i=1}^s d^2 z_i
|z_i|^{-4\rho j_1}|1-z_i|^{-4\rho j_2-2}
\prod_{i<j}|z_i-z_j|^{4\rho}
\eqn\dx
$$}

\def\dxi{$$
\left(\G(-j_2+m_2+s)\over \G(-j_2+m_2)\right)^2
=(-)^s\D(j_2-m_2+1)\D(j_3-m_3+1)
\eqn\dxi
$$}

\def\dxii{$$
e^{\a J^-_0}{\cal V}_{j\; m} e^{-\a J_0^-}=
\sum_{k=0}^\infty {\a^k\over k!}[J_0^-,{\cal V}_{j\; m}]_k
\eqn\dxii
$$}

\def\dxiia{$$
e^{\a J^-_0}{\cal V}_{j\; m} e^{-\a J_0^-}=
\sum_{k=0}^\infty {\a^k\over k!}
[J_0^-,{\cal V}_{j\; m}]_k,
\eqn\dxiia
$$}

\def\dxiib{$$
[{J_0}^-,{\cal V}_{j\; m}]_{0}=
{\cal V}_{j\; m}\qquad , \qquad [{J_0}^{-},{\cal V}_{j\;
m}]_{k}=[{J_0}^-,[{J_0}^-,{\cal V}_{j\; m}]_{k-1}].
\eqn\dxiib
$$}

\def\dxiic{$$
[J^-_0,{\cal V}_{j\; m}]=-(j+m)
{\cal V}_{j\; m-1}\qquad,\qquad
[J^-_0,{\cal V}_{j\; m}]_k=(-)^k
{\G(j+m+1)\over \G(j+m-k+1)}{\cal V}_{j\; m-k}.
\eqn\dxiic
$$}

\def\dxiii{$$
\sum_{k_2,k_3=0}^{\infty}
{\a^{k_2+k_3}\over k_2! k_3!}
{\Gamma(j_2+m_2+1) \over \Gamma(j_2+m_2+1-k_2)}
{\Gamma(j_3+m_3+1) \over \Gamma(j_3+m_3+1-k_3)}
{\cal
A}^{j_1\;j_2\;\;\;\;\;\;\;\;\;j_3}_{j_1\;m_2-k_2\;m_3-k_3}=
$$
$$
\sum_{k_1=0}^{\infty}{(-\a)^{k_1}\over k_1!}
{\Gamma(2j_1+1)\over \Gamma(2j_1+1-k_1)}
{\cal
A}^{j_1\;\;\;\;\;\;\;j_2\;\;\;j_3}_{j_1-k_1\;m_2\;m_3}.
\eqn\dxiii
$$}

\def\dxiv{$$
{\cal
A}^{j_1\;\;\;\;\;\;\;j_2\;\;\;j_3}_{j_1-k_1\;m_2\;m_3}=
\sum_{n=0}^{k_1}
\left({\Gamma(-2j_1)\over \Gamma(k_1-2j_1)}
{\Gamma (k_1+1)\over \G(n+1)\Gamma(k_1+1-n)}
{\Gamma(j_2+m_2+1)\over \Gamma(j_2+m_2+1-n)}
$$
$$
{\Gamma(j_3+m_3+1)\over \Gamma(j_3+m_3+1-k_1+n)}
{\G(j_2-m_2+n+s)\over \G(j_2-m_2+n)}\right)^2
I(j,k)
\eqn\dxiv
$$}

\def\dxvi{$$
{\cal A}^{j_1\;\;\;j_2\;\;\;j_3}_{m_1\;m_2\;m_3}
={\cal C}^2 {\cal I}(j_1,j_2,j_3,k),
\eqn\dxvi
$$}

\def\dxvib{$$
\eqalign{
{\cal
A}^{j_1\;\;\;\;\;\;\;j_2\;\;\;j_3}_{j_1-k_1\;m_2\;m_3}=
\Biggl(&
{\G(-2j_1)\over \G(k_1-2j_1)}
\sum_{n=0}^{k_1} {1\over n!}
{\G(k_1+1)\over \G(k_1+1-n)}
{\G(j_2+m_2+1)\over\G(j_2+m_2+1-n)}
\cr &
{\G(j_3+m_3+1)\over \G(j_3+m_3+1-k_1+n)}
{\G(-j_2+m_2-n+s)\over \G(-j_2+m_2-n)}
\Biggl)^2 {\cal I}(j_1,j_2,j_3,k)}
\eqn\dxvib
$$}

\def\dxvic{$$
{\cal C}=
{\G(j_1+m_1+1)\over \G(2j_1+1)}
{\G(j_3+m_3+1)\over \G(j_1-m_1)\G(-m_2-j_2)}
$$
$$
\sum_{n=0}^{k_1}
{\G(n+j_2-m_2+1)\G(j_1-m_1+n)\G(-j_2-m_2+n)
\over \G(n+1-j_1+j_3-m_2) \G(n-j_1-j_3-m_2)\G(n+1)}
\eqn\dxvic
$$}

\def\dxvii{$$
{\cal C}={\G(-2j_1)\over \G(-j_1-m_1)} {\G(j_3+m_3+1)\over
\G(-j_2+m_2)} {\G(j_1+j_3+m_2+1)\over \G(-j_1+j_3-m_2+1)}
$$
$$
\lim_{x\rightarrow 1}
{_3F_2}(j_2-m_2+1,m_1-j_1,-j_2-m_2;-j_1-j_3-m_2,-j_1
+j_3-m_2+1|x).
\eqn\dxvii
$$}

\def\dxix{$$\eqalign{
{\cal I}(j_1,j_2,j_3,k)&=
M^s \G(-s)\int \prod_{i=1}^s d^2 z_i |z_i|^{-4\rho j_1}
|1-z_i|^{-4\rho j_2-2}\prod_{i<j}|z_i-z_j|^{4\rho}\cr
&=\rho^{2s}(-\pi \D(1-\rho)M)^s\left( {\G(-2j_1+s)
\over \G(-2j_1)}\right)^2 {\cal Y}_{13}{\cal Y}_{2}\;\;,\cr}
\eqn\dxix
$$}

\def\dxx{$$
{\cal Y}_{13}=\prod_{i=0}^{s-1}\D(-2j_1\rho+i\rho)
\D(-2j_3\rho +i\rho)
\quad , \quad {\cal Y}_{2}=
\G(-s)\G(s+1)\prod_{i=0}^{s-1}\D((i+1)\rho)
\D(-2j_2\rho+i\rho).
\eqn\dxx
$$}

\def\dxxi{$$
2j_2+1=-{2\over 3}m_2 \qquad,\qquad
2j_i+1={2\over 3}m_i\quad {\rm for} \quad i=1,3.
\eqn\dxxi
$$}

\def\dxxia{$$
j_2={s\over 2}-{1\over 4}\qquad , \qquad m_2=
-{3\over 2}s-{3\over 4}.
\eqn\dxxia
$$}

\def\dxxib{$$
{\cal C}=-{1\over 2}{\G(-2j_1)\over \G(-2j_1+s)}
{\G(2j_1+{5\over 4})\over \G(2j_1+{5\over 4}-s)}
{\G({1\over 2})\G({1\over 4})\over
\G(-2s-{1\over 2})\G(s+{5\over 4})}.
\eqn\dxxib
$$}

\def\dxxic{$$
{\cal C}^2=\rho^{2s}\left({\G(-2j_1)\over
\G(-2j_1+s)}\right)^2{\G({1\over 4})\over \G({3\over
4})}\prod_{i=1}^3\D\Bigl(2j_i+{5\over 4}\Bigr).
\eqn\dxxic
$$}

\def\dxxii{$$
2j+1=\pm {2\over 3}m+\varepsilon r(j,\g),
\eqn\dxxii
$$}

\def\dxxiii{$$
\prod_{i=0}^{n-1}(i+x)={\G(n+x)\over \G(x)}
\qquad ,\qquad
\G(-n+\varepsilon)={(-)^n\over \varepsilon \G(n+1)}+
{\cal O}(1)\quad{\rm for}\quad
n\in {\rm I\!N}
\eqn\dxxiii
$$}

\def\dxxiv{$$
{\cal Y}_{13}=\rho^{2s-2}
\prod_{i=1,3}
\D(-8j_i-2)\D\Bigl(2j_i+{3\over 4}\Bigr).
\eqn\dxxiv
$$}

\def\dxxv{$$
s=2j_2+{1\over 2}+{\e\over 2}\widetilde{r}
\eqn\dxxv
$$}

\def\dxxvi{$$
{\cal Y}_2 =
(-)^{s+1}\rho^{2s+1}{\G({3\over 4})\over \G({1\over
4})}\D(-8j_2-2)\D\Bigl(2j_2+{3\over 4}\Bigr).
\eqn\dxxvi
$$}

\def\dxxvia{$$
{\cal R}={\G(s+{\widetilde{r}+4\over 8})\over
\G({\widetilde{r}+4\over 8})\G(s+1)},
\eqn\dxxvia
$$}

\def\dxxvii{$$
{\cal I}(j_1,j_2,j_3,k)=-\rho^{6s-1}M^s
(\pi \D(-\rho))^s\left({\G(-2j_1+s)\over
\G(-2j_1)}\right)^2\prod_{i=1}^3\D(-8j_i-2)
\D(2j_i+{3\over 4})
\eqn\dxxvii
$$}

\def\dxxviia{$$
{\cal A}(j_1,j_2,j_3)=
-\rho^{6s-1} (-\pi M\D(-\rho))^s\prod_{i=1}^3
\D(-8j_i-2)\D\Bigl(2j_i+{3\over 4}\Bigr)\D
\Bigl(2j_i+{5\over 4}\Bigr),
\eqn\dxxviia
$$}

\def\dxxviii{$$
{\cal A}(j_1,j_2,j_3)=
\widetilde{M}^s\prod_{i=1}^3\D(-4j_i-1)
\eqn\dxxviii
$$}

\def\dxxviiia{$$
{\cal V}_{p}^{\pm}=e^{(-\sqrt{2}\pm |p|)\phi +ipX}.
\eqn\dxxviiia
$$}

\def\dxxviiib{$$
{\cal A}(p_1,p_2,p_3)=
\widetilde{M}^{s}\prod_{i=1}^3\D(1-\sqrt{2}|p_i|)\qquad ,
\qquad
s=1+{1\over 2}\sum_{i=1}^3\left({|p_i|\over
\sqrt{2}}-1\right).
\eqn\dxxviiib
$$}

\def\dxxviiic{$$
A(p_1,p_2,p_3)
=[\mu \D(-\rho)]^s\prod_{i=1}^3 (-\pi)
\D(1-\sqrt{2}|p_i|)\quad,\quad
{s\over 2}=1+{1\over 2}\sum_{i=1}^3
\left({|p_i|\over \sqrt{2}}-1\right).
\eqn\dxxviiic
$$}

\def\gi{$$
{\langle {\cal V}_{j_2\; m_2}
{\cal V}_{j_3 \;m_3+1} \rangle \over
\langle {\cal V}_{j_2\; m_2+1}
{\cal V}_{j_3 \;m_3} \rangle} =
-{j_2-m_2\over j_3-m_3} =
-{j_3+m_3+1 \over j_2+m_2+1}.
\eqn\gi
$$}

\def\gii{$$
{\cal A}^{j_2\;\;\;j_3}_{m_2\;m_3}=
(-)^s\D(1+j_2-m_2)\D(1+j_3-m_3)I(j_2,j_3,\rho),
\eqn\gii
$$}

\def\giii{$$
I(j_2,j_3,k)=M^s \G(-s)\lim_{\varepsilon\rightarrow 0}
 \int \prod_{n=1}^s d^2 z_n
|z_n|^{-4\rho \varepsilon i}|1-z_n|^{-4\rho j_2-2}
\prod_{n<m}|z_n-z_m|^{4\rho}=
$$
$$
-(\pi M\D(-\rho))^s  \D(1-s)\D(\rho s)
\lim_{\varepsilon\rightarrow 0}
\D(1-2\rho\varepsilon i)\D((\varepsilon i-j_2+j_3)\rho))
\D((\varepsilon i+j_2-j_3)\rho).
\eqn\giii
$$}

\def\giv{$$
\delta(j_2-j_3)=\lim_{\varepsilon\rightarrow 0}
{1\over \pi}{\varepsilon \over \varepsilon^2+(j_2-j_3)^2}.
\eqn\giv
$$}

\def\gv{$$
{\cal I}(j_2,j_3,k)=2\pi i\rho\delta(j_2-j_3)
(\pi M\D(-\rho))^s  \D(1-s)\D(\rho s)
\eqn\gv
$$}

\def\gvi{$$
\left( {\G(-j_2+m_2+s)\over \G(-j_2+m_2)}\right)^2=
\D(1+j_2-m_2)\D (1+j_3-m_3)
\eqn\gvi
$$}

\def\gvii{$$
{\cal A}^{j_2\;\;\;j_3}_{m_2\;m_3}=
2\pi i\rho \delta(j_2-j_3)(-\pi M \D(-\rho))^s
 \D(1-s)\D(\rho s)
\D(1+j_2-m_2)\D(1+j_3-m_3)
\eqn\gvii
$$}

\def\gviiia{$$
\langle {\cal V}_{j_1\; m_1}(0)
{\cal V}_{j_2\; m_2}(1)\rangle=
\lim_{z_s\rightarrow \infty}\langle
{\cal V}_{j_1\; m_1}(0){\cal V}_{j_2\; m_2}(1)
\,\,\b(z_s){\bar \b}({\bar z}_s)e^{-{2\over \ap}
 \phi(z_s,{\bar z_s})}\,\,{\cal Q} ^{s-1}\,\rangle.
\eqn\gviiia
$$}

\def\gviiib{$$
s=j_1+j_2+1\qquad,\qquad m_1+m_2=0.
\eqn\gviiib
$$}

\def\gviii{$$
{\cal P}^{-1}{\p^s{\cal P}\over \p z_1
\dots \p z_s}
{\bar {\cal P}}^{-1}{\p^s{\bar {\cal P}}\over
 \p {\bar z}_1 \dots \p {\bar z}_s}=
(-)^s\,\D(1+j_1-m_1)\D(1+j_2-m_2)\prod_{i=1}^s
|z_i|^{-2}\,|1-z_i|^{-2},
\eqn\gviii
$$}

\def\gix{$$
{\cal P}=\prod_{i=1}^sz_i^{m_1-j_1}(1-z_i)^{m_2-j_2}
\prod_{i<j}^s(z_i-z_j).
\eqn\gix
$$}

\def\gx{$$
\langle {\cal V}_{j_1\; m_1}(0){\cal V}_{j_2\;
m_2}(1)\rangle=
(-)^s M^s\G(-s)\D(1+j_1-m_1)\D(1+j_2-m_2)
$$
$$
\int \prod_{i=1}^{s-1} d^2z_i|z_i|^{-4\rho j_1-2}
|1-z_i|^{-4\rho j_2-2}\prod_{i<j}^{s-1}
|z_i-z_j|^{4\rho}.
\eqn\gx
$$}

\def\gxa{$$
\langle {\cal V}_{j_1\; m_1}(0){\cal V}_{j_2\; m_2}
(1)\rangle=(-)^s
M^s\G(-s)\D(1+j_1-m_1)\D(1+j_2-m_2)\G(s)
$$
$$
(\pi\D(1-\rho))^{s-1}
\prod_{i=1}^{s-1}\D(i\rho)\prod_{i=0}^{s-2}
\D(-2j_1\rho+i\rho) \D(-2j_2\rho+i\rho)\D(1+\rho(s-i)).
\eqn\gxa
$$}

\def\gxi{$$
\langle {\cal V}_{j\; m}(0){\cal V}_{j\; -m}(1)\rangle=
(-\pi M\D(-\rho))^s\D(1+j-m)\D(1+j+m)s\D(1-s)\D(\rho s),
\eqn\gxi
$$}

\def\gxii{$$
\langle {\cal V}_{j\; m}(0){\cal V}_{j\; -m}(1)\rangle=
{\widetilde{M}^{2j+1}\over 2j+1}(\D(-4j-1))^2,
\eqn\gxii
$$}

\def\gxiii{$$
\langle {\cal V}_p^+(0){\cal V}_{-p}^+(1)\rangle=
{\widetilde{M}^{|p|\over \sqrt{2}}\over \sqrt{2}|p|}
(\D(1-\sqrt{2}|p|))^2,
\eqn\gxiii
$$}

\def\gxiv{$$
\langle V_p^+(0)V_{-p}^+(1)\rangle=
{\mu^{\sqrt{2}|p|}\over \sqrt{2}|p|}
(\D(1-\sqrt{2}|p|))^2.
\eqn\gxiv
$$}

\def\gxv{$$
\langle V_p^+(0)V_{-p}^+(1)\rangle
\sim {M^{|p|\over \sqrt{2}}\over \sqrt{2}|p|}\D
\left(1-{|p|\over \sqrt{2}}\right)^2.
\eqn\gxv
$$}

\def\gxvi{$$
{\cal V}_p^+\rightarrow {{\cal V}_p^+ \over \D
\bigl({1\over 2}-{|p|\over \sqrt{2}}\bigr)}
\eqn\gxvi
$$}

\def\hi{$$
m_1+m_2+m_3=j_1+j_2+j_3=0
\eqno{\rm (A.1)}
$$}

\def\hii{$$
{\cal A}^{j_1\;\;\;j_2\;\;\;j_3}_{m_1\;m_2\;m_3}=
\int d^2 z\;\langle \,{\cal V}_{j_1\; m_1}(0)
{\cal V}_{j_2\; m_2}(1)
{\cal V}_{j_3\; m_3}(\infty)\,\,\b(z){\bar \b}
({\bar z})\, e^{-{2\over \ap}\,\phi(z,{\bar z}) }
\,\,\rangle_{M=0}
$$
$$
=M\G(-s)\int d^2 z\;\left( {j_1-m_1\over z}-
{j_2-m_2 \over 1-z}\right)\left({j_1-m_1\over
{\bar z}}-{j_2-m_2\over 1-{\bar z}}\right)
|z|^{{-4\rho }j_1}\, |1-z|^{{-4\rho}j_2}.
\eqno{\rm (A.2)}
$$}

\def\hiii{$$
{\cal A}^{j_1\;\;\;j_2\;\;\;j_3}_{m_1\;m_2\;m_3}
=M\G(-s)\left( {-m_1j_2+m_2 j_1\over j_1}\right)^2\int d^2 z
 |z|^{{-4\rho}j_1} |1-z|^{{-4\rho} j_2-2}+{\cal B}(j_1,j_2)
+{\cal B}^\ast (j_1,j_2)
\eqno{\rm (A.3)}
$$}

\def\hiiia{$$
{\cal B}(j_1,j_2)\sim \int d^2z {\p \over \p{\bar z}} \Bigl(
|z|^{-4 \rho j_1}|1-z|^{-4\rho j_2-2}(1-{\bar z})\Bigr).
\eqno{\rm (A.4)}
$$}

\def\hiiib{$$
\int_{\Sigma}d^2z{\p\over \p{\bar z}}(f(z,{\bar z}))=
-{i\over 2}\int_{\p \Sigma} dz f .
\eqno{\rm (A.5)}
$$}

\def\hiiic{$$
{\cal B}(j_1,j_2)\sim-{i\over 2}\,\
lim_{\e\rightarrow 0} \oint_\e dz\,
|z|^{-4\rho j_1}\,|1-z|^{-4\rho j_2 -2}\,
(1-{\bar z})=\pi\lim_{\e\rightarrow 0}
\e^{-8\,\rho j_2},
\eqno{\rm (A.6)}
$$}

\def\dxviii{$$
{\cal C}=(-)^s{\G(-2j_1)\G(s+1)\over \G(-2j_1+s)}
\sum_{i=0}^s{\G(m_1-j_1+i)\over \G(m_1-j_1)}
$$
$$
{\G(-m_2-j_2+i)\over\G(-m_2-j_2)}
{\G(-m_1-j_1+s-i)\over \G(-m_1-j_1)}
{\G(m_2-j_2+s-i)\over \G(m_2-j_2)}
{(-)^i\over (s-i)!i!}
\eqno{\rm (A.7)}
$$}

\def\hva{$$
{\cal C}_{s=1}={j_1m_2-m_1j_2\over j_1}
\eqno{\rm (A.8)}
$$}

\def\hvi{$$
{\cal A}^{j_1\;\;\;j_2\;\;\;j_3}_{m_1\;m_2\;m_3}
=\lim_{\varepsilon\rightarrow 0}{\cal
A}^{j_1\;\;\;j_2\;\;\;j_3}_{m_1\;m_2\;m_3}=
-{9\over 4}\widetilde{M}\D(-1-4j_1)\D(-1-4j_3)\G(-s)
\lim_{\varepsilon\rightarrow 0}\D(-2\rho j_2)\D(\rho)
\eqn\hvi
$$}

\def\hvii{$$
\lim_{\e\rightarrow 0}\D(-2\rho
j_2)\D(\rho)\G(-s)=\G(-2j_2-1/2)
\lim_{\e\rightarrow 0}\D(2+(-8+2A)\e)\D(-4+16\e)
\eqn\hvii
$$}

\def\ki{$$
\Gamma(1+z-n)=(-1)^n{\Gamma(1+z)\Gamma(-z)
\over \Gamma(n-z)}\qquad, \qquad n\in {\rm I\!N},
\eqno{\rm (B.1)}
$$}

\def\kia{$$
\prod_{i=0}^{n-1}(i+x)={\G(n+x)\over \G(x)}
\;\;,\;\;
\D(x)\D(-x)=-{1\over x^2}\;\;,\;\;
\D(x)\D(1-x)=1\;\;,\;\;\D(1+x)=-x^2\D(x)
\eqno{\rm (B.2)}
$$}

\def\kiii{$$
\G(2x)={2^{2x-1}\over \sqrt{\pi}}\G(x)\G
\Bigl(x+{1\over 2}\Bigr)\qquad,\qquad
\D(2x)=2^{4x-1}\D(x)\D\Bigl(x+{1\over 2}\Bigr).
\eqno{\rm (B.3)}
$$}

\def\kii{$$
\lim _{\e\rightarrow 0}\G(-n+\e)={(-)^n\over
\e\G(n+1)}+{\cal O}(1)\quad {\rm for}\quad n\in {\rm I\!N}.
\eqno{\rm (B.4)}
$$}

\def\dxvia{$$
{_3F_2}(\a,\b,\gamma;\rho,\sigma|x)=
{\G(\rho)\G(\sigma)\over \G(\a)\G(\b)\G(\g)}
\sum_{\nu=0}^\infty{\G(\a+\nu)\G(\b+\nu)
\G(\g+\nu)\over
\G(\rho+\nu)\G(\sigma+\nu)}
{x^{\nu}\over \nu!}.
\eqno{\rm (B.5)}
$$}

\def\kv{$$
_3F_2(a,b,c;1+a-b,1+a-c|1)=
{\G(1+{a\over 2})\G(1+a-b)\G(1+a-c)
\G(1-b-c+{a\over 2})
\over
\G(1+a)\G(1-b+{a\over 2})\G(1-c+{a\over 2})
\G(1+a-b-c)}.
\eqno{\rm (B.6)}
$$}

\def\kvi{$$
{1\over m!}\int \prod_{i=1}^m\Bigl({1\over 2}
i dz_i d{\bar z_i}\Bigr)
\prod_{i=1}^m|z_i|^{2\a}|1-z_i|^{2\b}
\prod_{i<j}^m|z_i-z_j|^{4\rho}=
\pi^m\bigl(\D(1-\rho)\bigr)^m
$$
$$
\prod_{i=1}^m\D(i\rho)\prod_{i=0}^{m-1}
\D(1+\a+i\rho)\D(1+\b+i\rho)
\D(-1-\a-\b-(m-1+i)\rho).
\eqno{\rm (B.7)}
$$}

\chapter{INTRODUCTION}
One of the most interesting recent problems in string theory
is the study of the two-dimensional black hole solution. So
far, we have not been able to solve the many conceptual
problems of four-dimensional black holes, such as the
breakdown of quantum coherence at the endpoint of black hole
evaporation as advocated by Hawking \REF\haw{\rhaw}[\haw],
the existence of no-hair theorems or the cosmic censorship
hypothesis of Penrose.

With the hope that string theory may provide a natural
framework to address these problems in a simplified context,
Witten \REF\Wit{\rWit} [\Wit] proposed that the exact
conformal field theory that describes a black hole in
two-dimensional space-time can be formulated as an {\slU}
gauged Wess-Zumino-Witten (WZW) model. He argued that the
standard $c=1$ matrix model could be regarded as an analog
of the extreme Reissner-Nordstr\"om black hole of
four-dimensional general relativity, and that the $c=1$
$S$-matrix includes the formation and evaporation of black
holes in intermediate processes. However the exact
solubility of this model and the existence of the $W_\infty$
symmetry may constrain the physical events \REF\Witt{\rWitt}
[\Witt]. Ellis et al. \REF\emn{\remn} [\emn] concluded that
quantum coherence would be maintained during the black hole
evaporation due to these infinite number of conserved
quantities.

It is therefore important to understand the relation between
the 2-D black hole and the standard two-dimensional
non-critical string theory, which may further shed some
light on other important questions such as the background
independence in 2-D string theory.

The exact background metric of Witten's black hole has been
determined by Dijkgraaf, E. Verlinde and H. Verlinde
\REF\dvv{\rdvv} [\dvv]. They compared this model with $c=1$
coupled to Liouville and concluded that there might be
differences, since for example the reflection coefficients
of tachyons off the black hole geometry do not agree. Eguchi
et al \REF\eky{\reky} [\eky] have shown that the free field
BRST cohomologies of both models coincide, although Distler
and Nelson \REF\dn{\rdn} [\dn] emphasized that if one works
directly in the current algebra module there appear new
discrete states, which have no counterpart in $c=1$.
Independently Chaudhuri and Lykken \REF\cl{\rcl}[\cl] showed
that the discrete states of the black hole form a
$W_\infty$-type algebra, which has been analysed by Oz and
Marcus in detail \REF\mo{\rmo}[\mo].

The connection between the 2-D black hole and the $c=1$
standard non-critical string has been studied by Bershadsky
and Kutasov \REF\bk{\rbk}[\bk]. They used an equivalent
formulation of the {\slU} coset in terms of the Wakimoto
free field representation of \sl, which makes the
calculation of the $S$-matrix possible. They discussed the
spectrum of the 2D black hole and found that the bulk
correlation functions of tachyons (without interactions)
agree with the $c=1$ correlators without cosmological
constant. The purpose of this paper is to analyse whether
this equivalence still holds if the interactions of both
models are taken into account. We will do this at the level
of two- and three-point functions and find a remarkable
analogy. We will speculate whether this equivalence holds
for the $N$-point function, although concrete calculations
will appear in a later publication.

This paper is organized as follows. In section 2 we
summarize the {\slU} coset model following closely [\dvv].
We review the form of the currents, the gauged fixed action
and tachyon vertex operators in the Wakimoto free field
representation [\bk].
Furthermore, we review some facts about the cohomology of
the black hole [\eky] [\dn]. In section 3 we calculate the
three-point function of tachyon vertex operators in the
black hole background. We begin with the correlation
function containing one highest weight state and can obtain
the general three-point function using the {\sl} structure.
In section 4 we determine the form of the two-point function
of (not necessary) on shell tachyons. We compare our results
with $c=1$ non-critical string theory, the work of Dijkgraaf
et. al. [\dvv] and the deformed matrix model of Jevicki and
Yoneya \REF\jy{\rjy} [\jy], which has been proposed to be
the discrete matrix model formulation of the black hole
conformal field theory. In section 5 we give our conclusions
and some general remarks about $N$-point functions. In
appendix A we calculate the three-point function with one
screening as an illustrative example. We analyse contact
term interactions that arise in our computations in detail.
We review the most important formulas for our calculations
in appendix B.

\chapter{THE {\slU} WZW-MODEL}
In this section we review the Lagrangian formulation of the
{\slU} coset in terms of the Wakimoto  free field
representation and some basic facts about the cohomology of
the 2D black hole.

\sec{The Gauged WZW-Model}
The conformal field theory that describes a black hole in
two-dimensional target space-time has a Lagrangian
formulation in terms of a gauged WZW-model based on the
non-compact group {\sl} [\Wit]. The ungauged \sl-WZW model
has the following action:

\Ai

\noindent where $\Sigma$ is a Riemann surface with metric
tensor $h^{ij}$, $g:\Sigma \rightarrow${\sl} is an {\sl}
valued field on $\Sigma$ and $k$ is a real and positive
number. The Wess-Zumino term, which guarantees conformal
invariance is usually represented as:

\Aii

\noindent where $B$ is a three-dimensional manifold with
boundary $\Sigma$.

The action {\Ai} possesses a global {\sl$\times$\sl} symmetry,
since it is invariant under $g\rightarrow a g b^{-1}$ with
$a,b\in $\sl. The gauge-invariant generalization of {\Ai} is
formulated in terms of an Abelian gauge field $A$:

\Aiii

\noindent where $G=\left(\matrix{0&-i&\cr i&0&\cr}\right)$
for the Euclidean theory. If we impose the Lorentz gauge
condition $\p_{\alpha}A^{\alpha}=0$, the gauge slice can be
parametrized as $A^{\alpha}=\varepsilon^{\alpha
\beta}\p_{\beta}\Phi$, and the complete gauge-fixed action
of the Euclidean theory is given by [\dvv]:

\Aiv

\noindent where $\Phi$ is a free scalar field:

\Av

\noindent and $(b,c)$ is a spin $(1,0)$ system of ghosts:

\Avi

{}From {\Aiv} we see that the gauged WZW model can be
expressed through the ungauged \sl WZW model and the action
of the free fields $\Phi$, $b$ and $c$. As already remarked
in [\dvv] this makes the quantization straightforward.

To quantize the ungauged theory we notice that the \sl
symmetry gives rise to the conserved currents:

\Avii

\noindent where $t^1=i\sigma_1/2$, $t^2=\sigma_2/2$ and
$t^3=i\sigma_3/2$, and $\sigma_i$ are the Pauli matrices.
The modes of these currents satisfy the \sl current algebra
of level $k$, which is equivalent to the following operator
product expansion (OPE):

\Aviii

\sec{The Wakimoto Free Field Representation of \sl}

Like in the case of SU(2) \REF\d{\rd} [\d] the \sl current
algebra can be represented in terms of a spin $(1,0)$ chiral
bosonic $\beta$-$\gamma$ system and a free boson $\Phi$
satisfying \REF\w{\rw} [\w]:

\Bi

\noindent and the same holds for the antiholomorphic fields
${\bar \g}$ and ${\bar \b}$. If we introduce the notation
$\ap^2=2k-4$, the currents that satisfy the OPE  {\Aviii}
have the following form:

\Bii

The stress tensor follows from the Sugawara construction and
is given by the following expression:
\Biii

\noindent which after inserting the currents leads us to:

\Biv

\noindent This energy momentum tensor has a central charge
as a function of the level $k$ of the Kac-Moody algebra:
\Bv

To determine the form of the primary fields in the Wakimoto
representation, we need to introduce some basic facts about
the representation theory of \sl\foot{We thank L.
Alvarez-Gaum\'e for his notes on representation theory of
\sl.}. The basic fields from which we can build all the
other states are the Kac-Moody primaries
that satisfy:

\Bvii
and are characterized by the zero mode Casimir eigenvalue
$j$ and by the eigenvalue of $J^3_0$:

\Bvi

\noindent where $\Delta_0=-{1\over
2}(J_0^+J_0^-+J_0^-J_0^+)+(J_0^3)^2$. We can construct the
representation by acting with raising and lowering operators
$J_0^+$ and $J_0^-$. A solution of {\Bvi} is

\Bviii

\noindent If $j\in \R$ it is standard in the representation
theory of {\sl}
\REF\vil{\rvil}[\vil] to introduce new states $|j,m)$ that
satisfy

\Bxv

\noindent These recursion relations are satisfied up to a
function depending only on $j$, if we normalize the states
as:

\Bxva

\vskip 0.5cm

\noindent We will work with the states that satisfy {\Bviii}
unless otherwise stated. These Kac-Moody primaries define an
irreducible representation of {\sl} on which we can impose two
types of constraints [\dn]:

\item{\it 1.} {\it Hermiticity constraints.} We demand that
$\Delta_0$ and $J^3_0$ should have real eigenvalues. This
means $m\in\R$ and $j=-{1\over 2}+i\lambda $ or $j\in {\rm
I\!R}$ \REF\bih{\rbih}.
The types of Hermitian representations of \sl can be
classified as follows [\bih]:

\item{\bullet} Principal discrete series: Highest-weight or
lowest-weight representation. Contain a state annihilated by
${J_0}^+$ and ${J_0}^-$ respectively. They are one-sided and
infinite dimensional. From {\Bviii} we see that these
modules satisfy either $(j+m)$ or $(j-m)$ is an integer and:
\Bix
If in addition $2j$ is an integer, the representation is
double sided.

\item{\bullet} Principal continuous series. Satisfies
$j=-{1\over 2}+i\lambda$ with $\lambda$, $m\in \R$.

\item{\bullet} Supplementary series. In this case $j\in \R$,
but neither $j+m$ nor $j-m$ is an integer.

\item{\it 2.}{\it Unitarity constraints.} The states are
constrained to have positive norm, i.e. $J^+_0J^-_0$ and
$J^-_0J^+_0$ should have positive eigenvalues. This imposes
restrictions on the allowed values of $j$. We will not
impose any constraints of unitarity on our states.

The Kac-Moody primaries $|j,m\rangle$ are created, in the
Wakimoto representation, by the action of the following (not
normalized) ``tachyon'' vertex operator on the SL(2)-vacuum:

\Bxvi

\noindent and the same expression for the antiholomorphic
part. These vertex operators have the following OPE with the
energy momentum tensor and the currents:

\Bxvii

\vskip 0.5cm

\noindent Using the free field representation of the
currents {\Bii}, it is easy to check that ${\cal V}_{j\; m}$
satisfies the definition of a Kac-Moody primary {\Bvii}, as
well as {\Bvi} and {\Bviii}. The conformal dimension of
${\cal V}_{j\;m}$, obtained from the OPE with the energy
momentum tensor {\Biv}, is:

\Bxviii

The action associated with the energy momentum tensor {\Biv}
is:

\Bxix

If we would like to calculate correlation functions of the
vertex operators ${\cal V}_{j\; m}$ we would need a
screening charge, in order to guarantee for the charge
conservation arising from the zero mode integration.
This screening charge can be obtained directly from the WZW
model {\Ai}, if it is mapped with the Gauss-decomposition to
a Wakimoto free field theory \REF\gmom{\rgmom} [\gmom]. It
can be constructed independently [\bk] if we take into
account that the ungauged model has the \sl symmetry, which
leads to Ward identities for the correlators:

\Bxx

\vskip 0.5cm

\noindent Since these identities should be satisfied in the
free field representation, the screening charge must have a
regular OPE with the stress tensor and the currents. We must
also take into account that only positive integer powers of
$\b$ are well defined through bosonization, as we will see
in a moment.
The screening that satisfies these conditions can be
represented as the following surface integral:

\Bxxi

\noindent One can easily check the identities [\d]

\Bxxii

\vskip 0.5cm

\noindent The total derivative appearing in the last OPE
requires a careful treatment of contact terms\foot{We thank
G. Moore for pointing this out to us.}, as we will do later.
This comes from the fact that we are working with a surface
integral and not with a contour integral, where this
contribution generically vanishes. As is known to be correct
for a Coulomb gas model \REF\dfa{\rdfa}[\dfa] or Liouville
theory \REF\gl{\rgl}[\gl] \REF\dfk{\rdfk}[\dfk]
\REF\dtwo{\rdtwo}[\dtwo], the screening charge will be added
to the action and considered as the interaction of the model
[\bk]:

\Bxxiii

\vskip 0.5cm

The parameter $M$ has been identified with the mass of the
black hole [\bk] [\eky] that is responsible for the
non-trivial background of the theory.

To evaluate correlation functions, it will be useful to
bosonize the $\b$-$\g$ system as follows
\REF\fms{\rfms}[\fms]:
\Bxxiv
\noindent where $u$ and $v$ are ordinary bosons

\Bxxv

\vskip 0.3cm

\noindent This allows us to make sense of fractional and
negative powers of $\g$, while $\b^{\alpha}$ is well defined
only for $\alpha$ positive integer.

As already emphasized in [\bk] the realization of the sigma
model {\Ai} with a $\b$-$\g$ system, where the left and
right scheme are completely decoupled, corresponds to a
particular choice of contact terms. Different choices of
contact terms correspond to different choices of coordinates
in the coupling constant space
\REF\kutasov{\rkutasov}[\kutasov]
\REF\moore{\rmoore}[\moore], i.e. the space of metrics.

\sec{The Cohomology of the Gauged Model}

To construct the conformal field theory of the Euclidean
black hole we are interested in the coset \slU. To gauge the
U(1) subgroup, we introduce a gauge boson $X$ and a pair of
fermionic ghosts $B, C$ of spin $1, 0$ respectively:

\Ci

\noindent For the Euclidean theory the boson $X$ is compact
with radius $R=\sqrt{k}$ in units of the self dual radius.
The nilpotent BRST-charge of this symmetry is:

\Cii

\noindent The tachyon states of the gauged theory are the
dressed ghost number zero primary fields, which are
invariant under $Q^{U(1)}$\foot{Since $j$ and $m$ are
arbitrary at this moment, these operators will be called
tachyons.}:

\Ciii

\noindent The conformal weight of these fields is

\Civ

\noindent obtained from the OPE with the energy momentum
tensor of the U(1) gauged theory:

\Cv

\noindent The central charge of the gauged model:

\Cvi

\noindent is $c=26$ for $k=9/4$. The total anomaly cancels
if we take into account the pair of fermionic diffeomorphism
ghosts $b$, $c$ of spin $2$, $-1$, respectively, with
central charge $c_{gh}=-26$ and energy momentum tensor:

\Cvii

\noindent The total stress tensor is given by
$T_{total}=T_{\rm SL(2,\R)/U(1)}+T_{gh}$. For $k=9/4$ the
additional bosonic Liouville field decouples. We will see
later however, that it will be important in order to define
a consistent regularization of the correlation functions.

The BRST operator coming from the diffeomorphism invariance

\Cviii

\noindent satisfies
\Cix

\noindent Invariance under $Q^{Diff}$ up to a total
derivative implies the on shell condition

\Cixa

\noindent which is, for $k=9/4$:
\Cixb
Since we are dealing with two BRST charges, one can consider
the cohomology of $Q^{total}=Q^{U(1)}+Q^{Diff}$ or the
iterated cohomology, where the states are annihilated by
each of the BRST operators [\dn], as done in [\bk]. The
complete cohomology of the gauged WZW model has been
computed in [\dn] [\eky].

Using the Wakimoto free field representation (without
screening charge), Eguchi et al. [\eky] have determined the
cohomology of $Q^{total}$, which is believed to agree with
the iterated cohomology. It coincides with the cohomology of
$c=1$ Liouville theory without cosmological constant
\REF\lz{\rlz}[\lz]. From physical reasons one could expect
that (at least for tachyons) a similar correspondence may
exist if the interactions are taken into account
\REF\gm{\rgm} [\gm], although to our knowledge this has not
yet been proven.

Distler and Nelson [\dn] calculated both cohomologies
directly in the current algebra module. In addition to the
tachyon and $c=1$ discrete states, they obtained new
discrete states, which have no counterpart in Liouville
theory coupled to $c=1$ matter. The discrete states
classified in [\dn] are:

\Cx

\vskip 0.5cm

\noindent where $r,s$ are positive integers. The states
$\widetilde{\cal D}^{\pm}$ are the new discrete states,
while ${\cal D}^{\mp}$ appear in $c=1$ Liouville theory as
well as ${\cal C}$ and belong to the discrete and
supplementary series of {\sl} respectively. We will see later
how part of this spectrum appears as poles in tachyon
correlation functions, as it is known to hold for minimal
models or $c=1$ matter coupled to gravity [\gl] [\dfk]
[\dtwo].

As a prescription to ``glue'' together the left and right
representation, we will consider in analogy to [\dvv] only
spinless {\sl} primaries i.e. with $m={\bar m}$.
For the compact boson $X$ we have furthermore the
restriction:

\Cxvi
\vskip 0.5cm

\noindent States with $n_1=0$ are called winding modes. The
momentum modes which satisfy $n_2=0$ are out of the theory,
since they have spin.

\chapter{THE THREE-POINT FUNCTION}

In this section we calculate the three-point function of
tachyons in the black hole geometry. We begin with the
correlator containing one highest weight and will see that
the \sl structure fixes the correlators involving three
generic tachyons.

\sec{The Conservation Laws for $N$-point correlators}

The free field approach is appropriate to compute the
amplitudes, which obey a special energy sum rule, where the
number of screenings $s$ is a positive integer. As in the
case of the tachyon background, the desired correlators are
determined indirectly by an analytic continuation in $s$
[\gl] [\dfk] [\dtwo].
In general we wish to calculate the following scattering
amplitude:

\dia

\vskip 0.5cm

\noindent containing $N$ tachyons ${\cal V}_{j\;m}$ and $s$
screening charges. We will use a shorthand notation for the
vertex operators, where the antiholomorphic dependence with
$m={\bar m}$ will be understood throughout.

The conservation laws for an $N$-point amplitude are
obtained from the zero mode integration of the fields [\bk].
Integrating over the zero mode  $\phi_0=\phi-{\widetilde
\phi}$ \`a la Goulian and Li [\gl] gives:

\diii
where
\di

\noindent and we have absorbed a factor $-\ap/2$ into the
definition of the path integral. From the zero mode of $X$
it follows that
\dib

\noindent The number of zero modes of the $\b$-$\g$ system,
determined by the Riemann-Roch theorem, leads for spherical
topologies to the condition:

\dii

This constraint is equivalent to {\di} and {\dib} for states
of the form ${\Ciii}$.

For an arbitrary correlator the number of screenings will
not be a positive integer. We will determine the correlator
on the r.h.s. of {\diii} for integer screenings, and more
general amplitudes are obtained by an analytic continuation
in $s$. We begin with the three-point functions, since we
can use similar tricks to calculate two-point functions in a
simple way.

\sec{The Three-Point Function with One Highest-Weight State}

It turns out that the simplest way to address a general
three-point tachyon amplitude is to take one state belonging
to the discrete representation of \sl.
We choose in this section one of the tachyons as a
highest-weight state, for example $j_1=m_1$. We will see
later, that an arbitrary three-point function, where no
restriction to the representation to which the tachyons
belong is made, can be expressed as a function of this one.
The reason for this is that the {\sl} Clebsch-Gordan
coefficients can be analytically continued from one
representation of {\sl} to the others [\bih]. For the time
being, the level of the Kac-Moody algebra is still arbitrary
and we will need this restriction only when we want to make
the analytic continuation in the number of screenings at the
end.

Using SL(2,\IC) transformations on the integrand, we can fix
the three-tachyon vertex operators at
$(z_1,z_2,z_3)=(0,1,\infty)$\foot{We will drop the zero mode
in the next formulas.}:

\hskip 0.5cm

\div

\noindent This correlator has the following form after
bosonizing the $\b$-$\g$ system:

\dvi
\noindent where:
\dv

\noindent The derivatives of the above expression come from
the bosonization of $\b$ {\Bxxiv}. We can show that the
following identity holds:

\dvii

\noindent The proof uses the definition of the Vandermonde:

\dviii

\noindent where the sum goes over all permutations of the
indices. Inserting this expression for the Vandermonde in
the definition of ${\cal P}$ lead after a simple
calculation, to {\dvii}. We introduce the notation
$\D(x)=\G(x)/\G(1-x)$ and $\rho=-2/\ap^2=-1/(k-2)$. The
complete expression for the amplitude is:

\dix

\noindent We have used the identity:

\dxi

\noindent which holds for integer $s$. The remaining
integral:

\dx

\noindent can be solved using the Dotsenko-Fateev (B.9)
formula of [\dfa].
A careful treatment of the regularization for the case
$k=9/4$ is needed. We will discuss this solution and the
analytic continuation to arbitrary $s$ later. From the above
simple result, we can already see that \sl fusion rules
appear (see for example the appendix of [\cl]). If the
second tachyon is also in the highest-weight module, i.e.
$m_2=j_2-{\rm I\!N}$, then the result vanishes unless the
conjugate of $j_3$ is also in the discrete representation
i.e. $M=J-{\rm I\!N}$, where $(J,M)=(-1-j_3,-m_3)$. We will
now see how this generalizes for an arbitrary three-point
correlator, where a proportionality to a Clebsch-Gordan
coefficient appears.

\sec{Three-Point Function of Arbitrary Tachyons}

After getting an expression for the three-point function
containing one highest-weight state and two generic
tachyons, we would like to see how we can obtain the
amplitude of three generic tachyons.  Acting with the
lowering operator $J_0^-=\oint J^-(z) dz$ we compute the
amplitude $ {\bf {\cal
A}}^{j_1\;\;\;\;\;\;\;j_2\;\;j_3}_{j_1-k_1\; m_2\;m_3}$,
where the holomorphic $m_1=j_1-k_1$ dependence has been
changed by an integer $k_1$.
We will make an analytic continuation in $k_1$ to
non-integer values, while for the time being $s$ will be
taken as an integer. Our computation shows that the general
three-point function of (not necessarily on shell) tachyons
has the form:

\dxvi

\noindent where ${\cal C}$ is essentially the {\sl}
Clebsch-Gordan coefficient, whose expression we now
calculate. The square takes the antiholomorphic ${\bar m}$
dependence into account. We will use the
Baker-Campbell-Hausdorff formula:

\dxiia

\noindent where we have defined

\dxiib

\noindent The lowering operator $J_0^-$ acts on the
holomorphic part of the vertex operators:

\dxiic

\noindent We are going to use the fact that $J_0^-$ commutes
with the screening charge ${\cal Q}$, which is actually only
true up to a total derivative. The
surface terms that appear are discussed in appendix A. It is
shown that they can be neglected in a particular region of
$(j_i,m_i)$, and the other regions can be obtained by
analytic continuation \REF\gs{\rgs}[\gs]. With the above
formulas and the fact that $J_0^-$ annihilates the vacuum,
we get an identity for the general amplitude as a function
of the one with one highest weight state, if we identify in
powers of $\a$ the r.h.s. and l.h.s. of:

\dxiii

\vskip .5cm

\noindent Taking into account the antiholomorphic ${\bar m}$
dependence and formula {\dix} we get:

\dxvib

\vskip 0.5cm

\noindent Our aim is to give up the condition that $k_1$ is
an integer. First we notice that the above sum can be
extended to $\infty$ and written in terms of the generalized
hypergeometric function $_3F_2$ (B.5), which has a
definition in terms of the Pochhammer double-loop contour
integral that possesses a unique analytic continuation to
the whole complex plane of all its indices [\bih]. Comparing
{\dxvi} and {\dxvib}:

\dxvii

\vskip 0.5cm

\noindent The appearance of the generalized hypergeometric
function in our result is natural, since this function is
always present in the theory of \sl Clebsch-Gordan
coefficients [\bih]. It can be expanded in $s$, which will
be useful to check the analytic continuation in $k_1$ in
simple examples, as we do in appendix A. In general $_3F_2$
has a complicated expansion as a sum of $\G$ functions.
Fortunately, for on shell tachyons, which is the case we are
interested in, we get a simple result.

We will consider three tachyons on the right branch
satisfying $j\geq -1/2$.
To fulfil the $m$ conservation law {\dib}, we take without
loss of generality $m_2\leq 0$. If we choose $k=9/4$ the on
shell conditions are {\Cixb}:

\dxxi

\noindent This fixes $j_2$ as a function of the screening,

\dxxia

\noindent With this kinematic relations it is easy to check
that the generalized hypergeometric function is well-poised
and we can apply Dixon's theorem \REF\eder{\reder}[\eder] to
simplify $_3F_2$. With (B.6) it is straightforward to
obtain:

\dxxic

\hskip 0.40cm

\noindent We now have to evaluate the remaining part of the
amplitude {\dxvi}.
The integral ${\cal I}(j_i,k)$ can be solved using the
Dotsenko-Fateev (B.9) formula [\dfa]:

\dxix

\noindent where
\dxx

\vskip 0.5cm

\noindent This result holds for arbitrary level $k$ but
integer screenings, so that it has to be transformed in
order to obtain an expression valid for a non-integer $s$.
This analytic continuation will be done \`a la Di Francesco
and Kutasov [\dfk], using the on shell condition and taking
the kinematics into account. From the above definition we
notice that ${\cal Y}_2$ has dangerous singularities for
$k=9/4$, i.e. $\rho=-4$, while ${\cal Y}_{13}$ is well
defined since $j_1$ and $j_3$ could be chosen arbitrarily.

We first consider ${\cal Y}_{13}$. Choosing $j_1,j_3\notin
\IZ /8$ and the kinematic relations for $\rho=-4$, we
obtain:

\dxxiv

\vskip 0.5cm

The product ${\cal Y}_2$ is more subtle because we have to
find a regularization that preserves the symmetries of the
theory (see \REF\ms{\rms} [\ms] for related issues to this
subject). We will shift the level $k$ away from the critical
value and introduce a small parameter $\e\rightarrow 0$,
that can be set to zero for ${\cal Y}_{13}$. This can be
done by setting $\rho=-4+16\e$, i.e. $k=9/4+\e$, and taking
the limit $\e\rightarrow 0$. With this modification of the
level, the total central charge will be of order
$\varepsilon$ and the Liouville field $e^{\g \varphi_L}$ has
to be taken into account in order to get an anomaly-free
theory. For the on shell condition we make the general
ansatz:

\dxxii

\noindent which in the limit $\varepsilon \rightarrow 0$
reduces to our previous result.
Here $r(j,\g)$ could be, in principle, an arbitrary function
of the $j$'s and the Liouville dressing $\g$. With the
kinematics {\dxxii}, we get:
\dxxv

\noindent with ${\widetilde
r}=r(j_1,\g_1)-r(j_2,\g_2)+r(j_3,\g_3)$.
After simple transformations we obtain:

\vskip 0.2cm

\dxxvi

\vskip 0.5cm

\noindent In the above formula there appears a
multiplicative factor:

\dxxvia

\vskip 0.3cm

\noindent
which comes from the regularization. We have no further
constraint on the parameter $\widetilde{r}$ that appears in
the above expression. The best we can do is to fix it by
physical arguments. Choosing $\widetilde{r}=4$ will imply
that the three-point function factorizes in leg factors and
the four-point function will have obvious symmetry
properties. This imposes strong constraints on
$\widetilde{r}$. We will have to set $\widetilde{r}=4$ and
the contribution of the renormalization factor is one. We
will come back to this point when we discuss the $N$-point
functions in our next publication.

Our result for the on shell tachyon three-point amplitude is
obtained from {\dxvi}, {\dxix}, {\dxxiv} and {\dxxvi}:
\dxxviia
\hskip 1cm

\noindent which can be transformed finally to\foot{We absorb
a factor of two in the definition of the path integral}:
\dxxviii

\noindent where $\widetilde{M}=-\pi M\D(-\rho)\rho^{-2}$.

We can translate our notation to the one used in $c=1$. It
has been shown
[\eky] [\bk] that the black hole mass operator corresponds
to a discrete state of $c=1$ up to BRST exact terms. In this
language  the $\b$-$\g$ contribution of the correlators
cancels out and the vertex operators can be written as:

\dxxviiia
\vskip 0.3cm

\noindent Here $\pm$ denotes the tachyon vertex operators on
the right (wrong) branch, which represent the incoming
(outgoing) wave at infinity \REF\eguchi{\re}[\eguchi]. We
can write the result for the three-point function of on
shell tachyons in the black hole background as:

\dxxviiib

\hskip 0.1cm

\noindent Which can be compared with the $c=1$ three-point
function, with tachyonic background [\dfk]:
\dxxviiic

\noindent There appear several remarkable features in our
result:
\item{\bullet} In order to get finite correlation functions,
the black hole mass has been infinitely renormalized, as is
done for the cosmological constant $\mu$ in Liouville. Here
it is known to be equivalent to the replacing
$e^{-\sqrt{2}\varphi}\rightarrow \varphi
e^{-\sqrt{2}\varphi}$, which may have physical interesting
consequences. Perhaps this will be the case for the black
hole model as well.

\item{\bullet} From the zero mode integration of the
$N$-point function in both theories, we see that the number
of screenings for the black hole is twice the value of the
$c=1$ model.

\item{\bullet} The amplitude can be factorized in leg poles,
which (with this normalization) have resonance poles where
the $c=1$ discrete states are placed. The new discrete
states of Distler and Nelson do not appear. The explanation
of this is that these extra states are BRST trivial in the
Wakimoto representation. This has been shown by Bershadsky
and Kutasov for the first examples [\bk].

\chapter{THE TWO-POINT FUNCTION}

In this section we will calculate the two-point function and
compare our results with standard non-critical string
theory, the positions of the resonance poles that follow
from the quantum mechanical analysis carried out by
Dijkgraaf et. al. [\dvv] and the correlators of the deformed
matrix model of Jevicki and Yoneya [\jy]. If we have $c\leq
1$ matter coupled to Liouville theory the simplest way to
construct a two-point function of on shell tachyons is to
use the three-point function. One of the operators is then
set to be the dressed identity and this is the derivative
with respect to the cosmological constant $\mu$ of the
two-point function [\dfk]. In this case the situation is
different because the interaction is not a tachyon but a
discrete state of $c=1$ [\bk] [\eky]. Fortunately we can
construct the two-point function of (not necessarily on
shell) tachyons in the black hole background. We will do
that with two independent methods, as a double check of our
computation.

\sec{Two-point Function with $s-1$ Integrated Screenings}

We can perform a direct computation, fixing the position of
one of the screenings at $z_s=\infty$ and evaluating the
remaining $(s-1)$-integrals:

\gviiia

\vskip 0.5cm

\noindent We can follow closely the steps of the previous
computation, although in this case it will be much simpler.
We use the representation {\Bxvi} for the vertex operators
and the bosonization formulas for the $\b$-$\g$ system. The
zero mode integrations give:

\gviiib

\noindent Due to the kinematic relations satisfied by the
two-point amplitude, the part of the integrand coming from
the $\b$-$\g$ system can be written as:

\gviii

\noindent where

\gix

\noindent This can easily be seen with the substitution
$z_i=1/y_i$.
After evaluating the $\phi$ contractions and taking
$z_s\rightarrow \infty$,
the complete $1\rightarrow 1$ amplitude is reduced to the
evaluation of the following integral:

\gx

\vskip 0.3cm

\noindent The result is well defined for $\rho\neq -4$ and
can be obtained from (B.6):

\gxa

\vskip 0.5cm

\noindent In the next section we show that the two-point
function of (not necessarily) on shell tachyons is only
different from zero for $j_1=j_2=j$, so that we set $s=2j+1$
and, from the conservation law $m_1=-m_2=m\geq0$, we obtain
for an arbitrary level:

\gxi
\vskip 0.3cm
\noindent If $k\neq 9/4$ and the tachyons do not belong to a
discrete representation of {\sl}, the above amplitude has
one divergence, which appears for $s$ integer and comes from
the zero mode integration. For $k=9/4$ we demand $j\notin
\IZ/2$, which implies that $s$ is non-integer. This can
always be done, since the above expression is well defined
in this case. The final expression for on shell tachyons is:

\gxii

\vskip 0.5cm

\noindent where $\widetilde{M}$ is the renormalized black
hole mass previously defined. We will comment on this result
in section (4.3).

\sec{Two-Point Function with $s$ Integrated Screenings}

We can obtain the answer from the three-point tachyon
amplitude that contains one highest-weight state ${\cal
V}_{j_1\; j_1}$, taking the limit $j_1=i\e \rightarrow 0$.
In this way we are fixing only two points of the SL(2,\IC)
invariant $1\rightarrow 1$ amplitude. The result will
contain a divergence coming from the volume of the dilation
group \REF\sei{\rsei}[\sei] [\gm].

First we can show that this amplitude is diagonal by pushing
$J_0^+$ and $J_0^-$ through the correlator [\bih]. Here
again we will use that the screening charge commutes with
the currents.  We obtain the relations

\gi

\vskip 0.5cm

\noindent From the $X$ zero mode integration we get $m_2=m$
and $m_3=-1-m$, so that this equation has two solutions, one
with $j_2=-1-j_3$, which contains no screenings, and another
one with $j_2=j_3$, which is a two-point function with
$s=2j_2+1$ screenings. The first one is normalized to one up
to the divergence $\G(0)$ coming from the zero mode
integral. We now compute the two-point function with
screenings.

\noindent We first consider the case of arbitrary $k$ and
take the limit $k\rightarrow 9/4$ at the end of the
calculation. From {\dix}, we obtain  for $j_1=i\e$:

\gii

\noindent where
\giii

\vskip 0.5cm

\noindent Here we have simplified the products of well
defined $\D$-functions. To evaluate the limit we take into
account (B.4) and the following representation of the delta
distribution:
\giv

\vskip 0.3cm

\noindent In total we obtain for the two-point function of
generic (in general off shell) tachyons:

\gvii

\vskip 0.7cm

\noindent where $s=2j_2+1$ and the level is arbitrary. This
agrees with
{\gxi} up to a factor $s$ and the volume of the dilation
group $\delta(j_2-j_3)$.

\sec{Comparison to other Approaches}

Using the $c=1$ language, we obtain for the two-point
function {\gxii} of two Seiberg on shell tachyons in the
black hole background\foot{We have absorbed a factor of 2 as
before.},

\gxiii

\vskip 0.2cm

\noindent which we now compare with different models.

The two-point function of $c=1$ with cosmological constant
is:

\gxiv

\noindent Comparing with the two-point function in the black
hole background we find the same features as for the
three-point function. The pole structure is the same as the
one of the two-point function of $c=1$, while the screenings
differ by a factor of two.

The two-point function of the deformed matrix model is given
by the expression [\jy]:
\gxv

\noindent The position of the resonance poles are in
agreement with
the quantum mechanical analysis of Dijkgraaf et. al. [\dvv].
Here only half of the states of $c=1$ (the supplementary
series) appear as poles in the leg factors. Both two-point
functions can be reconciled if we take into account that we
can renormalize the tachyon vertex operators in a different
way according to the \sl representation theory. If in the
normalization {\Bxva} we use the on shell condition and take
into account the antiholomorphic piece, we renormalize our
operators as follows:

\gxvi

\noindent Here a regular function depending on $p$ has been
dropped, which is not determined by {\Bxva}. The two-point
function of these operators in the black hole background
agrees with {\gxv}. It is simple to see how the $N$-point
function of these differently normalized tachyons behaves.
The $N$-point function with chirality $(+,\dots,+,-)$, will
be divided by a factor $\D(-s-(N-1)/2)$, coming from the
state with the opposite chirality. This will imply, that
$(+,\dots,+,-)$ odd-point functions of these operators
vanish for positive integer screenings, if they were
previously finite. For even point functions this factor is
of course irrelevant for integer $s$ so that they are finite
in this case. Which one is the correct physical
normalization is to be clarified. For related problems see
[\gm].

We can compare the result that we obtain for this
three-point function with the deformed matrix model where
these odd-point functions generically vanish. While the zero
of our three-point function holds for integer screenings and
has it's origin in leg factors, this does not seem to be the
case for the deformed matrix model, where the
nonfactorizable part of the three-point function coming from
the collective field theory has zeros for arbitrary
odd-point functions [\jy]. We think it must be found a way
out of that if we would like to reconcile both approaches.

\chapter{CONCLUSIONS AND OUTLOOK}
Using the Wakimoto free field representation of the \slU
Euclidean black hole, we have found that tachyon two- and
three-point correlation functions share a remarkable analogy
with the amplitudes of $c=1$ coupled to Liouville at
non-vanishing cosmological constant. This observation was
made by Bershadsky and Kutasov for the bulk amplitudes, i.e.
those amplitudes where no screening charge is needed to
satisfy the total charge balance.
In order to have non-vanishing correlation functions, we
have infinitely renormalized the black hole mass. This is a
well-known phenomenon for $c=1$ and in this case it has
interesting physical consequences. Perhaps this is the case
for the black hole mass as well, further investigation is
desirable.

The amplitudes factorize in leg factors, which have poles at
all the discrete states of $c=1$. The new discrete states of
Distler and Nelson [\dn] do not appear, because they are
BRST trivial in the Wakimoto representation, as checked in
[\bk] for the first examples. With our renormalization of
the operators, we do not reproduce the pole structure of the
correlators of the deformed matrix model [\jy]. Whether this
discrepancy could be merely a normalization of the operators
was discussed. The scaling of the correlators is different
from $c=1$, but can be reproduced with the substitution
$\mu^2=M$.

Of course, there are a lot of interesting questions,
suggested by these observations. An important question is to
see whether this relation to $c=1$ persists for the
$N$-point function. Our preliminary analysis of the pole
structure indicates that again all the poles of $c=1$
appear.
Further work is needed to obtain a closed expression for all
$N$-point functions on the sphere. In particular, it will be
nice to see if we are able to obtain the result using the
Ward identities of $c=1$. We will report on that in a later
publication.

Furthermore it will be interesting to see whether it is
possible to find a relation to the model of Vafa and Mukhi
\REF\vm{\rvm} [\vm] and to use the powerful methods of
topological field theory to evaluate correlators on higher
genus in the continuum approach.

As an important application of the obtained two point
function we consider the determination of the exact Penrose
diagram for the quantum theory in the Minkowski case. Work
in this direction is in progress.

\vskip 2cm

\noindent {\bf Acknowledgements}

\noindent We would specially like to thank G. Moore and D.
Kutasov for many useful discussions and the reading of the
manuscript. We also acknowledge discussions with L.
Alvarez-Gaum\'e, J. Distler, V. Dotsenko, M. Douglas, W.
Lerche, E. Verlinde and C. Vafa.

\noindent K. Becker is supported by the Deutsche
Forschungsgemeinschaft and M. Becker is supported by a
scholarship Graduiertenf\"orderung des Landes NRW.

\vskip 1cm

\centerline{\bf APPENDIX A}

\noindent As an illustrative example we will consider in
more detail the three-point function with one screening.  In
this case we have:
\hi
\noindent and the correlator is given by:

\hii
\vskip 0.5cm
\noindent To evaluate this integral directly, without any
restriction on the $m$ dependence of the three vertex
operators, we have to use partial integration. We get:

\hiii
\vskip 0.5cm
\noindent where ${\cal B}(j_1,j_2)$ is the part coming from
the boundaries of the region of integration. It is
proportional to:
\hiiia
\vskip 0.3cm
\noindent This integral can be evaluated using (see appendix
A of [\moore]):

\hiiib
\vskip 0.3cm

\noindent We obtain:

\hiiic

\noindent where the integral is around a small circle of
radius $\e$ around 1. The contribution of the surface term
is zero for $j_2>0$, finite for $j_2=0$
and diverges for $j_2<0$. As argued by Green and
Seiberg [\gs], one has to add a finite contact term in the
case where the boundary terms are finite and an infinite
term if they diverge in order to render the amplitude
analytic. These contact terms can be avoided by calculating
amplitudes in an appropriate kinematic configuration, where
the contact terms are not needed, and then analytically
continuing to the desired kinematics.
For $s>1$, our argumentation will be the same as for $s=1$,
and we will restrict the values of $j$ to the regions where
the boundary terms vanish. This also means that we will
restrict to the kinematic regions where $J_0^-$ commutes
with the screening charge. We can compare this result with
the one, following from {\dxvi} which is based on the
analytic continuation in $k_1=j_1-m_1$ to non-integer
values. We obtain ${\cal C}$ for integer screenings
expanding {\dxvii}:

\dxviii

\noindent which for $s=1$ gives:

\hva

\noindent With {\dxvi} and {\dx} the amplitude becomes
(A.3), where ${\cal B}(j_1,j_2)=0$. With this explicit
example, one can already see that the analytic continuation
in $k_1$ is correct. The integral can be solved using (B.7)
for $m=1$ and the result is {\dxxviii}.

\vfill
\endpage

\centerline{\bf APPENDIX B}

\noindent For convenience we will collect the identities of
$\G$ functions that we have used in our computations:
\ki

\kia

\kiii

\noindent To regularize the result of the singular integrals
we use:

\kii

\noindent The definition of the hypergeometric function,
which we need in section 3 to compute the three-point
function of generic tachyons, is:

\dxvia

\noindent The following identity, known as ``Dixon's
theorem'', is useful to evaluate the three-point function of
on shell tachyons:

\kv

\vskip 0.5cm

\noindent Finally we have used the Dotsenko-Fateev (B.9)
formula to evaluate the integrals:

\kvi

\vskip 1cm

\refout
\end